\documentclass[a4paper,11pt]{article}

\pdfoutput=1

\usepackage{jcappub}
\usepackage{float}
\usepackage[T1]{fontenc}
\usepackage{bm}
\usepackage{siunitx}
\usepackage{subcaption}
\title{Finding the distribution of matter using lenses -- I: deconvolution-based reconstruction with CMB lensing}

\author[a,b]{Ajoy Dawn}

\author[c]{Jun-Qian Jiang}

\author[a, b, d]{Dhiraj Kumar Hazra}

\author[e]{Benjamin L'Huillier}

\author[c,f]{Arman Shafieloo}

\affiliation[a]{The Institute of Mathematical Sciences, CIT Campus, Chennai 600 113, India}
\affiliation[b]{Homi Bhabha National Institute, Training School Complex, Anushakti Nagar, Mumbai 400094, India}
\affiliation[c]{Korea Astronomy and Space Science Institute, Daejeon 34055, Korea}

\affiliation[d]{INAF/OAS Bologna, Osservatorio di Astrofisica e Scienza dello Spazio, Area della ricerca
CNR-INAF, via Gobetti 101, I-40129 Bologna, Italy}

\affiliation[e]{Department of Physics and Astronomy, Sejong University, Seoul 05006, Korea}

\affiliation[f]{University of Science and Technology, Daejeon 34113, Korea}

\emailAdd{ajoydawn@imsc.res.in}
\emailAdd{jiang@kasi.re.kr}
\emailAdd{dhiraj@imsc.res.in}
\emailAdd{benjamin@sejong.ac.kr}
\emailAdd{shafieloo@kasi.re.kr}

\abstract{
The matter power spectrum is one of the primary statistical descriptors of the large-scale distribution of matter in the Universe and provides a powerful probe of cosmic structure formation. Measurements of cosmic microwave background (CMB) lensing offer an integrated view of the matter distribution over a wide range of redshifts, enabling the reconstruction of the underlying matter power spectrum. In this work, we reconstruct the reference linear matter power spectrum $ P_\text{lin}(k,0)$ from the baseline joint CMB lensing measurements of Planck PR4, ACT DR6, and SPT-3G
using a covariance-weighted modified Richardson--Lucy(MRL) deconvolution algorithm. The reconstructed spectrum is found to be consistent with the fiducial linear prediction on large scales, while exhibiting a systematic enhancement for $k \gtrsim 0.1\,{\rm Mpc}^{-1}$, where nonlinear gravitational evolution becomes important. To investigate this behavior, we introduce a scale-dependent correction factor, $A
(k)$, defined through
$
P(k)=A(k)\,P_{\rm nl}(k),$
where $P_{\rm nl}(k)$ is the fiducial nonlinear matter power spectrum obtained from 2LPT
simulations. The reconstructed correction factor remains consistent with unity within $2\sigma$ confidence over the reconstructed range, indicating that the observed enhancement is well explained by the standard nonlinear evolution of the matter power spectrum. In addition, the reconstruction shows agreement with the fiducial BAO template around the  BAO feature at $k\sim(0.04-0.06)\ {\rm Mpc}^{-1}$, indicating that some BAO-scale information survives the lensing projection.
}

\begin{document}

\maketitle

\section{Introduction}
The matter power spectrum, $P_{\rm m}(k,z)$, is a central statistical
description of large-scale structure.  Its broadband shape, acoustic
oscillations, and nonlinear evolution encode complementary information about
the primordial perturbations, the matter content of the Universe, and the
growth of structure.  Modern observations usually constrain this spectrum
through projected two-point functions rather than by measuring
$P_{\rm m}(k,z)$ directly.  It is therefore useful to ask how much of the
scale dependence of the matter spectrum can be recovered from a given
projected observable without assuming a detailed parametric form for that
scale dependence.

CMB lensing is especially well suited for this question.  Gravitational
lensing remaps the primary CMB anisotropies and provides a direct probe of the
projected matter distribution between us and the last-scattering surface
\cite{Hu:2000ee,Hu:2001tn,Okamoto:2003zw,Lewis2006}.  The current generation
of CMB lensing measurements from \textit{Planck} PR4~\cite{Carron2022}, ACT
DR6~\cite{Qu2024ACT}, and SPT-3G M2PM~\cite{Ge2024SPT}, together with their
joint analysis~\cite{APS2026}, has reached percent-level sensitivity to the
lensing amplitude and provides the most precise CMB-lensing-only measurement
of structure growth to date.  These data are usually interpreted by fitting a
small number of cosmological parameters or an overall lensing amplitude.  The
same measured bandpowers and their window functions can also be viewed as an
inverse problem for the scale dependence of the matter spectrum.

Non-parametric and weakly parametric reconstructions of the matter power
spectrum have a long history.  Early work formulated CMB, galaxy clustering,
weak lensing, and cluster measurements as windowed measurements in
$k$-space~\cite{Tegmark:2002cy}, and lensing tomography was later developed as
a route to reconstructing three-dimensional matter and galaxy power spectra
\cite{Simon:2012dw}.  Cosmic-shear analyses have explored non-parametric
cosmology and direct matter-spectrum reconstruction
\cite{Taylor:2018mqm,Preston:2024ggf,Broxterman:2024oay}, with recent
KiDS-based deprojections finding sensitivity to nonlinear-scale suppression
\cite{Simon:2025xew,Broxterman:2025kpt}.  Closely related recent work has also
used DES Y3 3x2pt data with Planck CMB lensing to reconstruct
$P_{\rm m}(k,z)$ through a growth-factor expansion~\cite{Ye:2024rzp}, and has
used ACT CMB lensing together with DES cosmic shear, or multiple weak-lensing
surveys, to infer binned or free scale-dependent modifications of the nonlinear
matter spectrum~\cite{Sarmiento:2025yyh,Doux:2025vru}.  Recent applications of free-form primordial-spectrum reconstruction using Planck, ACT, and SPT data further demonstrate the utility of the Modified Richardson--Lucy framework~\cite{Chandra:2026byw}. Building on this
broader reconstruction program, we focus here on the information carried by the current CMB-lensing auto bandpowers themselves, using the combined
\textit{Planck}, ACT, and SPT measurements at the bandpower level.
Within the inversion framework based on the modified Richardson--Lucy deconvolution method, we study the recovery of a
scale-dependent reference matter spectrum, the survival of BAO-scale
oscillatory information, and the consistency of a fiducial nonlinear prescription.

We first ask whether the reconstruction can retain BAO-scale information
from the CMB lensing measurements. Acoustic oscillations are present in the
late-time matter power spectrum, but the line-of-sight lensing projection
smooths such features. Previous weak-lensing studies have shown that BAO
wiggles can be imprinted in projected lensing observables but are strongly
suppressed by projection and observational noise~\cite{Zhang:2008ta}
(see also~\cite{Baxter:2020qlr}), while forecasts for kinematic weak-lensing
tomography suggest that improved redshift information could enhance their
detectability~\cite{Ding:2019hmw}. We therefore use the BAO feature as a
test of whether any BAO-scale information survives the projection and
reconstruction.

We then investigate whether the reconstructed matter power spectrum shows
evidence for nonlinear evolution. In particular, we examine the enhancement
of the reconstructed spectrum relative to the fiducial linear prediction
at higher wavenumbers. Nonlinear structure formation, baryonic feedback,
and nonstandard dark-sector physics can all modify the matter power spectrum
on small scales~\cite{Amon:2022azi,Preston:2024ggf,Preston:2025tyl}. If the
observed enhancement is consistent with nonlinear evolution, we further
quantify any remaining deviation by introducing a redshift-independent,
scale-dependent correction factor $A(k)$ through
\begin{equation}
P(k)=A(k)P_{\rm nl}(k),
\end{equation}
where $P_{\rm nl}(k)$ is the fiducial nonlinear matter power spectrum.
Thus, the $A(k)$ reconstruction provides a quantitative consistency test
of the adopted nonlinear prescription at fixed background cosmology and
redshift evolution.

Our inversion method is based on the Richardson--Lucy deconvolution algorithm
\cite{Richardson1972,Lucy1974} and its modified cosmological implementation
developed for primordial-spectrum reconstruction
\cite{Shafieloo2004,Shafieloo2007}.  Modified Richardson--Lucy methods have
also been applied to simulated CMB-lensing spectra to reconstruct the
primordial power spectrum~\cite{Chandra:2021ydm}.  Here we adapt the modified Richardson--Lucy algorithm to observed \textit{Planck}+ACT+SPT CMB-lensing bandpowers, including the
bandpower window functions, and reconstruct either a reference linear
spectrum, $P_{\rm lin}(k,0)$, or the scale-dependent nonlinear modifier.
Here the $z=0$ linear spectrum is used as a reference spectrum for the line-of-sight projection.
Our reconstruction results follow the fiducial large-scale shape, show agreement with fiducial BAO template around $k\sim (0.04-0.06)\,{\rm Mpc}^{-1}$, and attribute the apparent high-$k$ excess in the linear reconstruction to non linear evolution.

This paper is organized as follows.  Section~2 presents the forward model,
bandpower compression, covariance-weighted MRL update, and Monte Carlo
uncertainty propagation.  Section~3 gives the validation tests and the
reconstruction from the joint CMB-lensing data, including the BAO extraction
and the nonlinear modifier analysis.  Section~4 discusses the interpretation
and limitations of the reconstruction, and summarizes our
conclusions.

\section{Methodology}

\subsection{Matter power spectrum parameterization}

Our objective is to reconstruct a reference linear matter power spectrum from CMB lensing observations. Since CMB lensing probes the matter distribution integrated over a broad range of redshifts, we do not reconstruct the matter power spectrum independently at each redshift. Instead, we adopt the present-day linear matter power spectrum $ P_\text{lin}(k,0)$ as a reference spectrum and assume linear, scale-independent growth to relate it to the matter power spectrum at higher redshifts entering the CMB lensing projection. 

Rather than treating the spectrum as a continuous unknown function, we represent it by its values at a finite set of nodes in wavenumber space. This discretization reduces the inverse problem to the reconstruction of a finite number of parameters while retaining sufficient flexibility to capture the scale dependence of the spectrum.

We define a strictly increasing grid of $N_k$ nodes,
\begin{equation}
k_1 < k_2 < \cdots < k_{N_k},
\end{equation}
and denote the value of the reconstructed spectrum at each node by
\begin{equation}
p_\nu \equiv P(k_\nu), \qquad \nu=1,\ldots,N_k.
\end{equation}

The spectrum at an arbitrary wavenumber is obtained through piecewise-linear interpolation in $\ln k$. Introducing

\begin{equation}
x \equiv \ln k,
\qquad
x_\nu \equiv \ln k_\nu,
\end{equation}
the interpolation is written as
\begin{equation}
P(k)
=
\sum_{\nu=1}^{N_k}
p_\nu\,S_\nu(\ln k),
\end{equation}
where $S_\nu(x)$ denotes the standard piecewise-linear basis (``hat'') functions,
\begin{equation}
S_\nu(x)=
\begin{cases}
\dfrac{x-x_{\nu-1}}{x_\nu-x_{\nu-1}},
&
\vspace{0.5cm}

x_{\nu-1}\le x \le x_\nu,
\\[1.2ex]
\dfrac{x_{\nu+1}-x}{x_{\nu+1}-x_\nu},
&
\vspace{0.5cm}

x_\nu\le x \le x_{\nu+1},
\\[1.2ex]
0,
&
\mathrm{otherwise},
\end{cases}
\end{equation}
with appropriate one-sided definitions at the endpoints. These basis functions satisfy
\begin{equation}
S_\mu(x_\nu)=\delta_{\mu\nu},
\qquad
\sum_{\nu=1}^{N_k}S_\nu(x)=1,
\end{equation}
ensuring that the interpolation exactly reproduces the nodal values while remaining continuous between adjacent nodes.

For the nonlinear reconstruction discussed in Section~\ref{sec:nonlinear_comparison}, we adopt an analogous parameterization for the scale-dependent correction factor,

\begin{equation}
A(k)
=
\sum_{\nu=1}^{N_k}
a_\nu\,S_\nu(\ln k),
\end{equation}

Here, $a_\nu\equiv A(k_\nu)$ denotes the value of the modifier evaluated at the $\nu$-th reconstruction nodes. The nonlinear matter power spectrum is then written as
\begin{equation}
P(k)=A(k)\,P_{\rm nl}(k),
\end{equation}
where $P_{\rm nl}(k)$ is the fiducial nonlinear reference spectrum. 
For the nonlinear analysis, the fiducial nonlinear matter power spectrum is generated using PySCo~\cite{Breton2023}  $N$-body simulations. We adopt PySCo primarily because it enables the
generation of a large number of computationally inexpensive mock
realizations required for the Monte Carlo analysis.
 Before adopting the
PySCo spectrum as the nonlinear reference, we verify that it agrees
reasonably well with the Mead2020~\cite{Mead2021} prescription over the
range of scales relevant to the reconstruction. The reconstruction grid
spans $2\times10^{-4}$--$0.7\,{\rm Mpc}^{-1}$, while the effective
data-constrained range is approximately $10^{-3}$--$0.7\,{\rm Mpc}^{-1}$,
with the main sensitivity extending to $k\sim0.6\,{\rm Mpc}^{-1}$.
Consequently, the PySCo nonlinear spectrum provides a suitable
simulation-based reference for the subsequent reconstruction.
\subsection{Forward model for CMB lensing}

To reconstruct the matter power spectrum from CMB lensing observations, a forward model is required that relates the underlying three-dimensional matter distribution to the observed lensing power spectrum. Throughout this work we assume a spatially flat Universe and adopt the Limber~\cite{LoVerde2008, Limber:1954zz} approximation,
which provides an accurate description of the CMB lensing power spectrum over the multipole range considered in our analysis.

Under these assumptions, the CMB convergence power spectrum is given by 
\begin{equation}
C_\ell^{\kappa\kappa}
=
\int_0^{z_*}
\mathrm{d}z\,
\frac{H(z)}{c\,\chi^2(z)}
W_{\rm CMB}^2(z)\,
P_\mathrm{m}\!\left(\frac{\ell+1/2}{\chi(z)},z\right),
\end{equation}
where $H(z)$ is the Hubble expansion rate, $\chi(z)$ is the comoving distance, and $z_*$ denotes the redshift of the last scattering surface. The CMB lensing kernel is
\begin{equation}
W_{\rm CMB}(z)
=
\frac{3\Omega_mH_0^2}
{2c\,H(z)}
(1+z)\chi(z)
\frac{\chi_*-\chi(z)}
{\chi_*},
\end{equation}
where $\chi_*$ is the comoving distance to the last scattering surface.

The observed CMB lensing measurements are reported in terms of the lensing potential power spectrum, which is related to the convergence spectrum through
\begin{equation}
C_\ell^{\phi\phi}
=
\frac{4}{[\ell(\ell+1)]^2}
C_\ell^{\kappa\kappa}.
\end{equation}

Combining the above expressions, the theoretical lensing potential power spectrum can be written as
\begin{equation}
C_\ell^{\phi\phi}
=
\frac{4}{[\ell(\ell+1)]^2}
\int_0^{z_*}
\mathrm{d}z\,
\frac{H(z)}
{c\,\chi^2(z)}
W_{\rm CMB}^2(z)\,
P_\mathrm{m}\!\left(\frac{\ell+1/2}{\chi(z)},z\right).
\label{eq:cl_phi}
\end{equation}

For the reconstruction of the reference linear matter power spectrum $P_\text{lin}(k,0)$, we assume scale-independent linear growth,
\begin{equation}
P_\mathrm{m}(k,z)
=
D_+^2(z)\,
P_{\rm lin}(k,0),
\label{eq:linear_growth}
\end{equation}
where $D_+(z)$ is the linear growth factor normalized to unity at the present epoch ($D_+(0) = 1$). This relation enables the matter power spectrum at any redshift to be expressed in terms of the present-day spectrum, which forms the quantity reconstructed in this work.

Substituting \ref{eq:linear_growth} in \ref{eq:cl_phi} we find, 
\begin{align}
C_\ell^{\phi\phi}
&=
\frac{4}{[\ell(\ell+1)]^2}
\int_0^{z_*}
\mathrm{d}z
\frac{H(z)}
{c\chi^2(z)}
W_{\rm CMB}^2(z)
D_+^2(z)
\sum_{\nu = 1}^{N_k}
p_\nu S_\nu
\left[
\ln\left(
\frac{\ell+1/2}{\chi(z)}
\right)
\right]
\nonumber\ \\
&=
\sum_{\nu=1}^{N_k}G_{\ell \nu}p_\nu,
\end{align}
with \begin{equation}
    G_{\ell \nu} = \frac{4}{[\ell(\ell+1)]^2}
\int_0^{z_*}
\mathrm{d}z\,  
\frac{H(z)}
{c\,\chi^2(z)}
W_{\rm CMB}^2(z)\,
D_+^2(z)S_\nu \left[ \ln \left(\frac{\ell+1/2 } {\chi(z)}\right)\right].
\end{equation}
This is the unbinned kernel mapping the node values $p_\nu = P_\text{lin}(k_\nu, 0)$ to the theory spectrum $ C_\ell^{\phi\phi}$.

\subsection{Bandpower compression and window functions}

The forward model described in the previous subsection provides the theoretical lensing potential power spectrum $C_\ell^{\phi\phi}$ at each multipole $\ell$. In practice, however, the CMB lensing measurements are not reported as individual multipoles but as binned bandpowers. Therefore, before comparing the theoretical prediction with the observations, the spectrum must be convolved with the experimental bandpower window functions provided by each collaboration.

For a given experiment $X\in\{\mathrm{Planck\ PR4},\,\mathrm{ACT\ DR6},\,\mathrm{SPT\text{-}3G\ M2PM\ }\}$, assuming $\mathcal{W}_{B\ell}^{X}$ denote the corresponding bandpower window matrix, where $B$ labels the observed bandpower. The theoretical prediction for the measured bandpowers can be expressed as,

\begin{equation}
t_B^{X}
=
\sum_\ell
\mathcal{W}_{B\ell}^{X}
C_\ell^{\phi\phi,\mathrm{th}}.
\end{equation}

Substituting the forward model into the above expression gives
\begin{equation}
t_B^{X}
=
\sum_\nu
G_{B\nu}^{X}\,
p_\nu,
\end{equation}
where the band-power kernel is defined as
\begin{equation}
G_{B\nu}^{X}
=
\sum_\ell
\mathcal{W}_{B\ell}^{X}
G_{\ell \nu}.
\end{equation}

Thus, for each experiment, the theoretical bandpower vector can be written compactly as
\begin{equation}
\mathbf t^{X}
=
\mathbf G^{X}\mathbf p,
\label{ref:forward_model_eq}
\end{equation}
where $\mathbf p$ contains the unknown values of the matter power spectrum at the reconstruction nodes.

To perform a joint reconstruction, the bandpower vectors from Planck PR4, ACT DR6, and SPT-3G M2PM are combined into a single data vector
\begin{equation}
\mathbf d
=
\begin{pmatrix}
\begin{array}{l}
\mathbf d^{\mathrm{Planck}}\\
\mathbf d^{\mathrm{ACT}}\\
\mathbf d^{\mathrm{SPT}}
\end{array}
\end{pmatrix},
\end{equation}
with the corresponding theoretical prediction
\begin{equation}
\mathbf t
=
\begin{pmatrix}
\begin{array}{l}
\mathbf t^{\mathrm{Planck}}\\
\mathbf t^{\mathrm{ACT}}\\
\mathbf t^{\mathrm{SPT}}
\end{array}
\end{pmatrix}
=
\mathbf G\,\mathbf p,
\end{equation}
where the combined kernel matrix is obtained by stacking the individual band-power kernels,
\begin{equation}
\mathbf G
=
\begin{pmatrix}
\begin{array}{l}
\mathbf G^{\mathrm{Planck}}\\
\mathbf G^{\mathrm{ACT}}\\
\mathbf G^{\mathrm{SPT}}
\end{array}
\end{pmatrix}.
\end{equation}

Following the joint lensing analysis, we approximate the covariance matrix by a block-diagonal form,
\begin{equation}
\mathbf C
\simeq
\begin{pmatrix}
\mathbf C^{\mathrm{Planck}} & 0 & 0\\
0 & \mathbf C^{\mathrm{ACT}} & 0\\
0 & 0 & \mathbf C^{\mathrm{SPT}}
\end{pmatrix},
\end{equation}
thereby neglecting cross-covariances between the three experiments. This approximation is justified by the limited overlap in sky coverage and the treatment adopted in the construction of the joint lensing likelihood.
The combined data vector, kernel matrix, and covariance matrix constitute the inputs to the modified Richardson--Lucy reconstruction described in the following subsection.
\subsection{Modified Richardson--Lucy reconstruction}

The forward model developed in the previous subsections provides a linear relation between the unknown matter power spectrum $(\mathbf{p} )$ and the predicted
CMB lensing bandpowers($\mathbf{t}$), as given in \ref{ref:forward_model_eq}.
Recovering $\mathbf{p}$ from the observed bandpowers therefore constitutes an inverse problem. Since the CMB lensing kernel projects information from a broad range of redshifts onto the observed multipoles, neighbouring $k$-modes contribute to the same bandpowers, rendering a direct inversion of $\mathbf{G}$ numerically unstable.

To solve this inverse problem, we employ the Modified Richardson--Lucy (MRL) deconvolution algorithm introduced by Shafieloo \& Souradeep~\cite{Shafieloo2004} and later extended in Ref.~\cite{Shafieloo2007}. The principal advantage of the Richardson--Lucy framework is that it naturally preserves the positivity of the reconstructed spectrum while iteratively improving the agreement between the theoretical prediction and the observed data. The original MRL algorithm was developed for reconstructing the primordial power spectrum from CMB temperature anisotropies. Here we adapt the same formalism to reconstruct the present-day matter power spectrum from CMB lensing bandpower measurements.

Before performing the reconstruction, the projection kernel is normalized column-wise,
\begin{equation}
\widetilde{G}_{B\nu}
=
\frac{G_{B\nu}}
{\sum_B G_{B\nu}},
\end{equation}
so that the contribution from each reconstruction node is normalized over all observed bandpowers.

Starting from an initial guess $\mathbf{p}^{0}$, the theoretical bandpowers at the $n$-th iteration are computed as
\begin{equation}
t_B^{n}
=
\sum_\nu
G_{B\nu}\,
p_\nu^{n}.
\end{equation}

The reconstructed spectrum is then updated according to
\begin{equation}
p_\nu^{n+1}
=
p_\nu^{n}
\left[
1
+
\tanh^2
\left(
(\mathbf d-\mathbf t^{n})^{\rm T}
\mathbf C^{-1}
(\mathbf d-\mathbf t^{n})
\right)
\sum_B
\widetilde{G}_{B\nu}
\frac{d_B-t_B^{n}}
{t_B^{n}}
\right],
\label{eq:MRL_update}
\end{equation}
where $\mathbf d$ denotes the observed bandpower vector, $\mathbf C$ is the covariance matrix of the measurements.

The summation term reduces to the standard Richardson--Lucy correction in the absence of the covariance-weighted damping factor, updating the reconstructed spectrum according to the fractional residual between the observed and predicted bandpowers. The second term acts as a damping factor that depends on the covariance-weighted goodness-of-fit,
\begin{equation}
\chi^{2}_{n}
=
(\mathbf d-\mathbf{ t^{n}})^{\rm T}
\mathbf C^{-1}
(\mathbf d-\mathbf {t^{n}}).
\end{equation}

As the reconstruction approaches the observed data, $\chi^{2}_n$  decreases and the damping term progressively suppresses further corrections, preventing the algorithm from fitting statistical fluctuations in the measurements. Compared with the original MRL formulation, which assumes independent uncertainties for each data point, the covariance-weighted implementation naturally accounts for the correlations between the measured band powers through the full covariance matrix.

The reconstruction is initialized from a positive-definite fiducial spectrum and updated iteratively. Throughout this work, ten iterations are adopted for all reconstructions, providing a stable solution while avoiding excessive sensitivity to statistical fluctuations in the data.
The same reconstruction framework is also employed in the nonlinear analysis presented later in this work. In that case, the reconstructed parameters correspond to the scale-dependent correction factor $A(k)$ rather than the linear matter power spectrum itself, while the forward model get modified and iterative reconstruction procedure remain unchanged.
\subsection{Monte Carlo uncertainty estimation}

The modified Richardson--Lucy algorithm provides a reconstructed quantity for a given realization of the observed CMB lensing bandpowers. To quantify the statistical uncertainty of the reconstruction, we perform a Monte Carlo analysis by propagating the observational covariance through the complete reconstruction pipeline.

Mock realizations of the observed bandpower vector are generated by drawing samples from a multivariate Gaussian distribution,

\begin{equation}
\mathbf d^{(r)}
\sim
\mathcal N
\left(
\mathbf d,
\mathbf C
\right),
\end{equation}

where $\mathbf d$ is the observed joint bandpower vector, $\mathbf C$ is the corresponding covariance matrix, and $r$ labels the individual realization. Each mock realization is reconstructed independently using the same forward model, reconstruction kernel, initial conditions, and convergence criteria adopted for the real data.

The ensemble of reconstructed spectra is then used to estimate the statistical properties of the reconstruction. At each reconstruction node $k_\nu$, the mean reconstructed spectrum is computed as
\begin{equation}
\overline{P}(k_\nu)
=
\frac{1}{N_{\rm mock}}
\sum_{r=1}^{N_{\rm mock}}
P_r(k_\nu),
\end{equation}
where, $N_{\rm mock}$ denotes the total number of Monte Carlo realizations. For our case, we have taken $N_{\rm mock} = 1000.$
The uncertainty bands presented throughout this work are obtained directly from the distribution of reconstructed spectra. We quote the central $68\%$ and $95\%$ confidence intervals using the 16th--84th and 2.5th--97.5th percentiles of the Monte Carlo ensemble, respectively. The same procedure is applied to all reconstructed quantities, including the linear matter power spectrum, the nonlinear correction factor $A(k)$, and the derived BAO ratio.
This Monte Carlo procedure propagates the observational uncertainties through the entire reconstruction pipeline without requiring linear error propagation, thereby providing a robust estimate of the uncertainty associated with the reconstructed quantities.

\subsection{Extraction of the BAO signal}

To isolate the oscillatory component of the reconstructed matter power spectrum, we divide the spectrum by a smooth reference spectrum obtained through Gaussian smoothing. Specifically, given a reconstructed spectrum $P(k)$ sampled on the reconstruction nodes, we construct a smooth spectrum,
\begin{equation}
P_{\rm smooth}(k)=\mathcal{G}_{\sigma}\left[P(k)\right],
\end{equation}
where $\mathcal{G}_{\sigma}$ denotes a Gaussian smoothing operator with smoothing width $\sigma$ applied in $\ln k$. The BAO signal is then defined by
\begin{equation}
R(k)=\frac{P(k)}{P_{\rm smooth}(k)}.
\end{equation}

The same procedure is applied to the fiducial linear matter power spectrum in order to construct the reference BAO template used for comparison.
The Gaussian smoothing scale is chosen to suppress the slowly varying broadband shape of the matter power spectrum while retaining the oscillatory BAO component. This smoothing procedure is intended solely to construct a smooth, no-wiggle reference spectrum for visualizing the oscillatory BAO component. It should not be confused with the peak-averaging or no-wiggle construction methods commonly used in BAO analyses, nor with density-field reconstruction techniques that aim to reverse nonlinear structure growth.
\section{Results}

\subsection{Zeroth-order consistency tests}

Before reconstructing the matter power spectrum from observational data, we test whether the reconstruction pipeline itself introduces spurious nonlinear features or artificial BAO oscillations.

\subsubsection{Consistency test with a linear input spectrum}

We first investigate whether the reconstruction algorithm introduces any artificial nonlinear features when the input spectrum is purely linear. To this end, mock CMB lensing spectra are generated from the fiducial linear matter power spectrum computed with CAMB~\cite{Lewis2000}, adopting the fiducial cosmological model used throughout this work, with Gaussian realizations drawn according to the corresponding covariance.
 The reconstruction pipeline is then applied to these mock realizations, and the reconstructed spectrum is subsequently compared with the fiducial input spectrum.

\begin{figure}[H]
    \centering

    \begin{minipage}{0.49\textwidth}
        \centering
        \includegraphics[width=\linewidth]{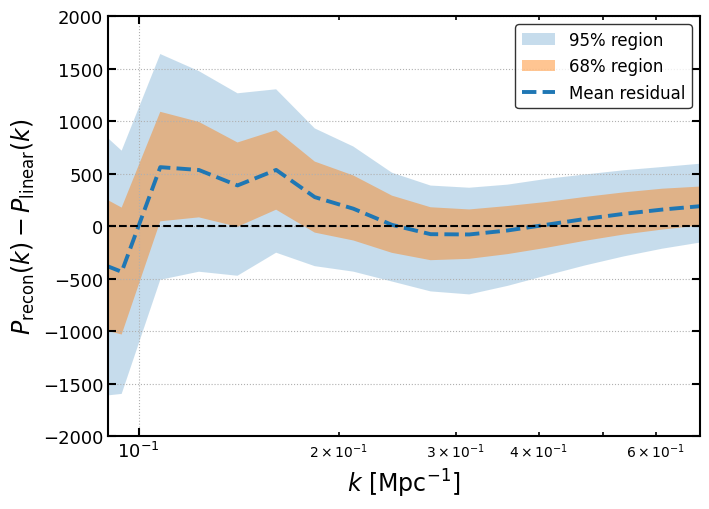}
    \end{minipage}
    \hfill
    \begin{minipage}{0.49\textwidth}
        \centering
        \includegraphics[width=\linewidth]{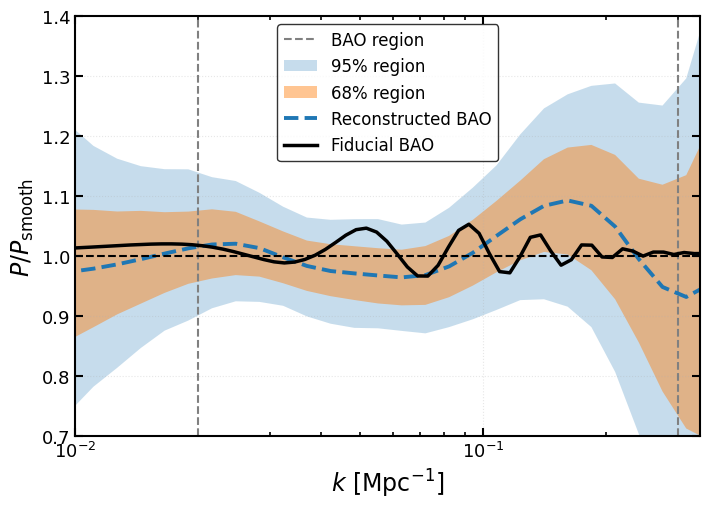}
    \end{minipage}

    \caption{
Left: Residuals between the reconstructed and fiducial linear matter power spectra for the linear-input consistency test. The dashed curve denotes the mean residual obtained from the Monte Carlo realizations.
Right: BAO null test using a smooth no-BAO input spectrum. The dashed
curve shows the mean reconstructed $P/P_{\rm smooth}$, while the solid
curve shows the fiducial BAO ratio for comparison. The absence of a
BAO-like oscillatory pattern in the reconstruction demonstrates that
the reconstruction pipeline does not generate spurious BAO features. In both panels, the orange and blue shaded regions correspond to the $68\%$ and $95\%$ confidence intervals, respectively.
}

    \label{fig:zeroth_order_tests}
\end{figure}

Figure~\ref{fig:zeroth_order_tests} left-hand panel shows the residual $P_\text{recon}(k) - P_\text{lin}(k)$ together with the $ 68\% $ and $ 95\%$ confidence regions obtained from Monte Carlo realizations. 
The residuals remain consistent with zero within  the $68\%$ ($1\sigma$) confidence region over the scales considered, indicating that the reconstruction procedure does not generate spurious nonlinear features by itself when the input spectrum is purely linear.

\subsubsection{Consistency test with a no-BAO input spectrum}

To further investigate whether the reconstruction algorithm generates artificial oscillatory features, we perform a BAO null test. We construct a smooth no-BAO matter power spectrum 
and use it to generate mock lensing data. The reconstruction is then performed following the same procedure as in the fiducial analysis.

Figure~\ref{fig:zeroth_order_tests}  right-hand panel shows the reconstruction of the smooth no-BAO input spectrum in terms of the ratio $P{(k)}/P_{\rm smooth}(k)$. The ratio remains consistent with unity within $68\%$ ($1\sigma$) confidence region over the scales considered, indicating that no statistically significant
BAO-like oscillatory features are generated when a smooth, no-BAO input
spectrum is reconstructed. This demonstrates that the reconstruction
procedure does not introduce spurious BAO features.

\subsection{Reconstruction under the linear matter-power-spectrum ansatz}

\begin{figure}[H]
    \centering
    \includegraphics[width=0.5\linewidth]{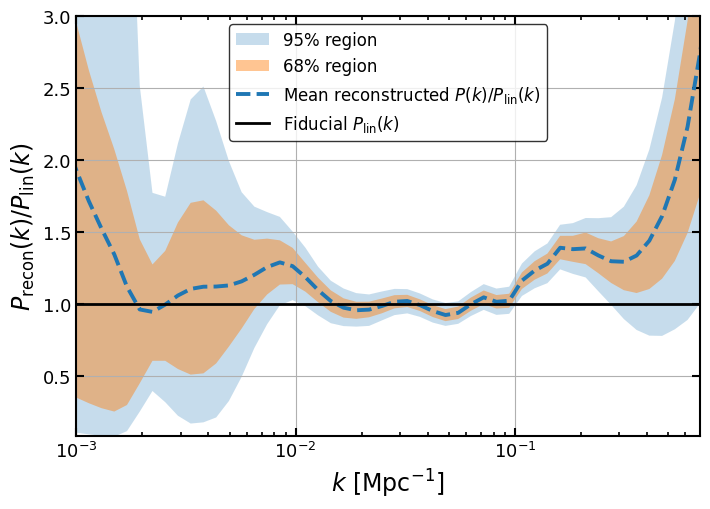}
    \caption{Ratio of the reconstructed matter power spectrum to the fiducial
    linear prediction, $P_{\rm recon}(k)/P_{\rm lin}(k)$, over the
    reconstructed wavenumber range. The dashed blue curve denotes the
    mean reconstruction, while the orange and blue shaded regions
    correspond to the $68\%$ and $95\%$ confidence intervals obtained
    from the Monte Carlo realizations. The black horizontal line
    indicates the fiducial linear expectation,
    $P_{\rm recon}/P_{\rm lin}=1$.
    }
    \label{fig:linear_reconstruction}
\end{figure}

Having established the consistency of the reconstruction pipeline, we now present the reconstructed linear matter power spectrum. Figure~\ref{fig:linear_reconstruction} shows the reconstructed spectrum relative to  the fiducial linear matter power spectrum corresponding to the best-fit $\Lambda$CDM cosmology adopted in the forward model. The blue dashed curve denotes the mean reconstruction obtained from the Monte Carlo ensemble, while the shaded regions represent the $68\%$ and $95\%$ confidence intervals propagated from the observational uncertainties.

The reconstruction is performed over the wavenumber range
$
10^{-3} \lesssim k \lesssim 0.7~{\rm Mpc}^{-1}$,
which corresponds to the scales where the joint CMB lensing measurements provide meaningful constraints on the matter power spectrum. Outside this interval, the reconstruction becomes progressively less constrained because of the broad projection kernel and the limited sensitivity of the measurements. Over most of the reconstructed range, the recovered spectrum closely follows the overall shape of the fiducial linear prediction. At $k\gtrsim0.1\,{\rm Mpc}^{-1}$, where nonlinear evolution is expected to become increasingly important, the reconstructed spectrum begins to show a systematic enhancement relative to the fiducial linear prediction. 

Figure~\ref{fig:linear_reconstruction} shows this behavior directly through the ratio $P_{\rm recon}(k)/P_{\rm lin}(k)$. The enhancement becomes statistically significant in the region where the reconstruction retains substantial sensitivity, around $0.1\lesssim k\lesssim0.2\,{\rm Mpc}^{-1}$, reaching more than $3\sigma$ level relative to the fiducial linear prediction. At still
higher wavenumbers, however, the sensitivity of the CMB lensing kernel
decreases and the uncertainty on the reconstruction grows, reducing the
significance of the apparent excess to approximately the $2\sigma$ level.
Thus, while the high-$k$ enhancement provides an indication that the
linear matter-power-spectrum ansatz is insufficient on smaller scales,
the reconstruction becomes progressively less constraining at the
highest wavenumbers. This motivates the comparison with nonlinear matter
power-spectrum models presented in the following subsection.

\subsection{Consistency with the BAO signature}

Beyond recovering the broadband shape of the matter power spectrum, an important question is whether the reconstruction is capable of preserving the oscillatory features associated with baryon acoustic oscillations (BAO). Since these oscillations represent only a small modulation of the underlying matter power spectrum, their recovery provides a stringent test of the reconstruction algorithm.

\begin{figure}[H]
    \centering

    \begin{subfigure}[t]{0.49\textwidth}
        \centering
        \includegraphics[width=\linewidth]{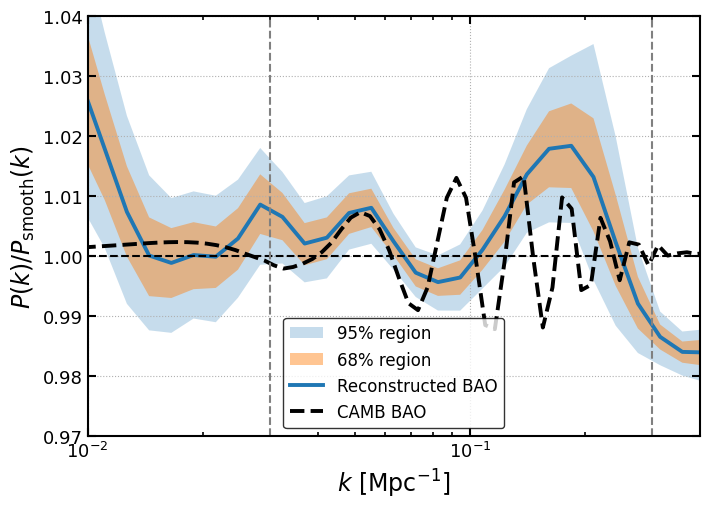}
        \caption{
        Comparison of the reconstructed BAO ratio $P_{\rm recon}(k)/P_{\rm smooth}(k)$ with the
corresponding fiducial BAO template. The reconstruction shows agreement
with the fiducial oscillatory pattern around
$k\sim\SI{0.05}{Mpc^{-1}}$.        }
        \label{fig:bao_reconstruction}
    \end{subfigure}
    \hfill
    \begin{subfigure}[t]{0.49\textwidth}
        \centering
        \includegraphics[width=\linewidth]{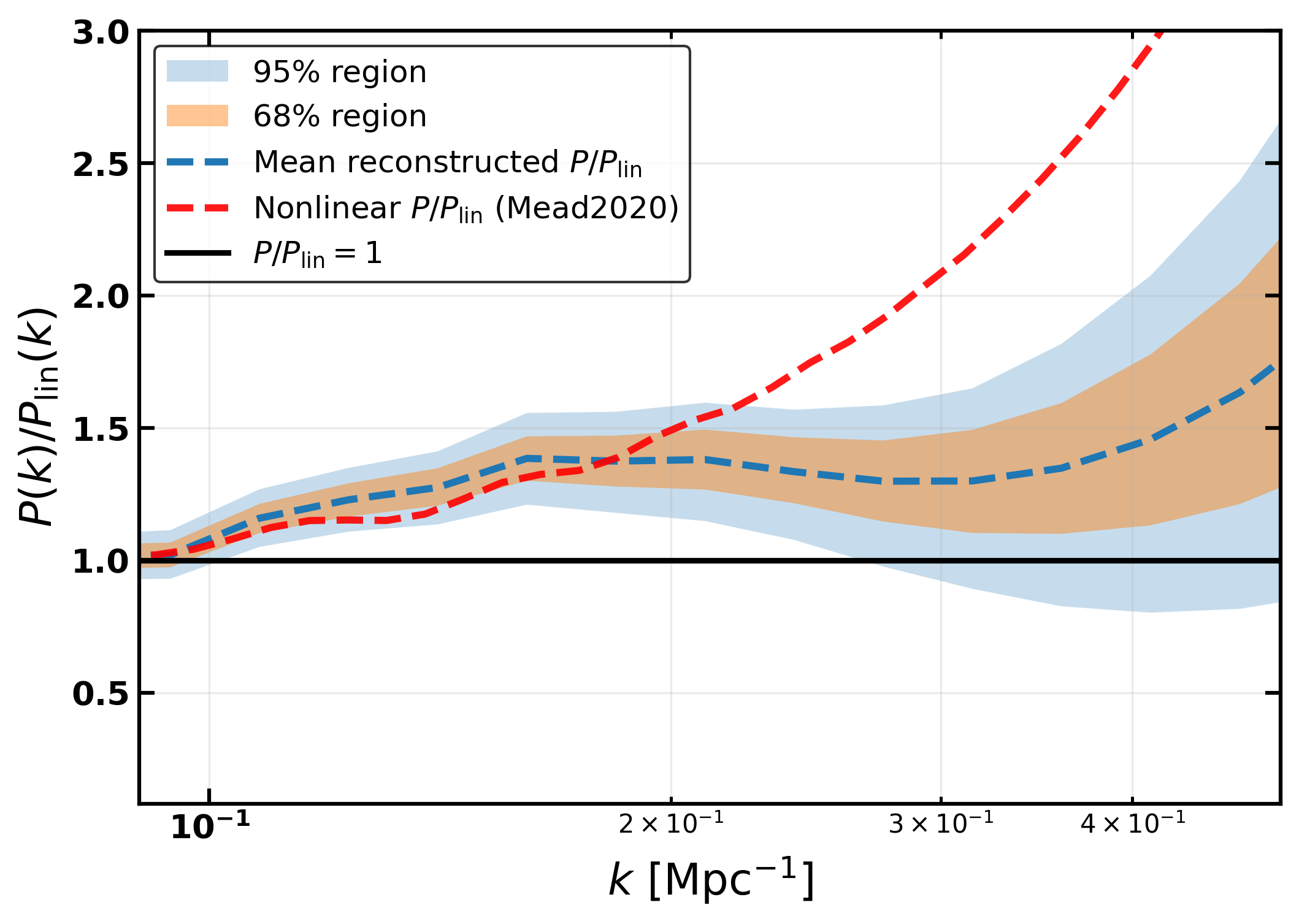}
        \caption{
        Ratio of the reconstructed matter power spectrum to the fiducial
        linear prediction, $P_{\rm recon}{(k)}/P_{\rm lin}(k)$. The dashed
        reconstructed ratio is compared with the corresponding nonlinear
        prediction, $P_{\rm nl}(k)/P_{\rm lin}(k)$, while the horizontal line
        indicates the fiducial linear expectation.
        }
        \label{fig:nonlinearity_check}
    \end{subfigure}

    \caption{
Tests of physical features in the matter power spectrum reconstructed from
the observed joint CMB lensing data under the linear matter-power-spectrum
ansatz.
(\subref{fig:bao_reconstruction}) comparison of the reconstructed
BAO ratio with the fiducial BAO template, testing the retention of BAO-scale information.
(\subref{fig:nonlinearity_check}) comparison of the reconstructed spectrum,
 expressed as
$P_{\rm recon}(k)/P_{\rm lin}(k)$, with the corresponding nonlinear
prediction, testing the interpretation of the high-$k$ enhancement.
}
    \label{fig:physical_interpretation}
\end{figure}

To isolate the oscillatory component, we divide both the reconstructed and fiducial matter power spectra by their corresponding smooth spectra, obtained using a Gaussian smoothing procedure. Figure~\ref{fig:bao_reconstruction} compares the reconstructed BAO ratio with the fiducial BAO template derived from the linear CAMB matter power spectrum.

Around $k\sim (0.04-0.06)\,{\rm Mpc}^{-1}$, the reconstructed oscillatory pattern follows the phase of the fiducial BAO template within the reconstructed uncertainties.
Outside the BAO region, the reconstructed ratio approaches the expected smooth behavior with no statistically significant oscillatory features. This is consistent with the limited sensitivity of the CMB lensing kernel at very low and very high wavenumbers, where the reconstruction is primarily constrained by the broadband shape of the matter power spectrum.

This agreement around $k\sim (0.04-0.06)\,{\rm Mpc}^{-1}$ indicates that some
BAO-scale information is retained by the modified Richardson--Lucy
reconstruction, although the full oscillatory BAO pattern is not
robustly recovered over the reconstructed range. This provides additional confidence that the reconstructed spectrum captures physically meaningful structure rather than numerical artifacts, and motivates the subsequent investigation of the nonlinear enhancement observed at smaller scales.

\subsection{Comparison with the nonlinear matter power spectrum}

The reconstruction presented in the previous subsection exhibits a systematic enhancement relative to the fiducial linear matter power spectrum, particularly towards smaller scales. Since nonlinear gravitational evolution is expected to increase the matter clustering on these scales, we now compare the reconstruction with the corresponding nonlinear matter power spectrum.

Figure~\ref{fig:nonlinearity_check} compares the ratio
$P_{\rm recon}(k)/P_{\rm lin}(k)$ with the corresponding nonlinear
prediction $P_{\rm nl}(k)/P_{\rm lin}(k)$. The nonlinear prediction shown here is the Mead2020 prescription evaluated at $z=0$. The reconstructed ratio follows the nonlinear prediction closely up to $k\sim0.2\,{\rm Mpc}^{-1}$, where the enhancement relative to the linear prediction is clearly captured.
 At higher wavenumbers, the reconstruction
departs increasingly from the nonlinear prediction as the reconstruction
becomes less constrained. This indicates that the systematic enhancement
observed in the previous subsection is consistent with nonlinear growth
over the range where the reconstruction retains sufficient sensitivity.

To quantify these residual deviations in a model-independent manner, we introduce a scale-dependent correction factor, $A(k)$, multiplying the fiducial nonlinear matter power spectrum. Since the PySCo nonlinear spectrum was found to be consistent with the
Mead2020 prediction over the relevant range, we use the PySCo spectrum
as the simulation-based reference for the subsequent reconstruction
of $A(k)$.The reconstruction of this correction factor is presented in the following subsection.
\subsection{Reconstruction of the correction factor
\texorpdfstring{$A(k)$}{A(k)}}
\label{sec:nonlinear_comparison}
The comparison presented in the previous subsection demonstrates that the reconstructed matter power spectrum is substantially better described by the nonlinear prediction than by the fiducial linear spectrum. To quantify any remaining deviation in a model-independent manner, we reconstruct the scale-dependent correction factor $A(k)$ introduced in Section~2, which multiplies the fiducial nonlinear matter power spectrum. The redshift dependence of the nonlinear matter power spectrum is obtained directly from the PySCo simulations, thereby incorporating the redshift evolution of nonlinear structure formation into the forward model.

\begin{figure}[H]
    \centering
    \includegraphics[width=0.65\linewidth]{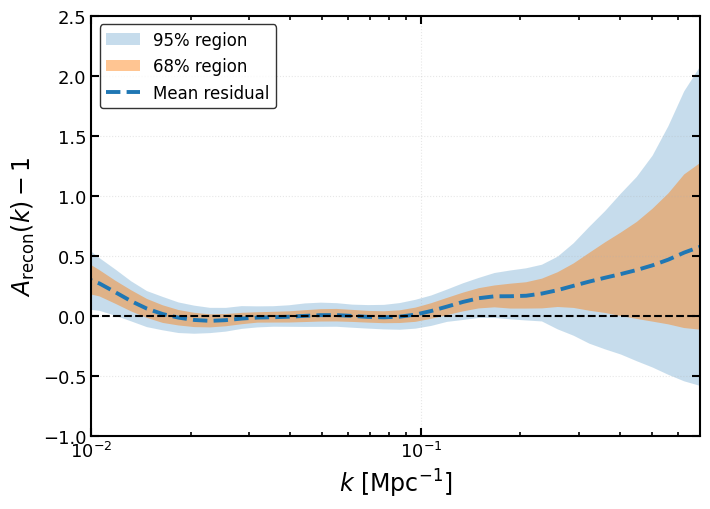}

    \caption{
    Reconstruction of the scale-dependent correction factor $A(k)$
    relative to the fiducial expectation $A(k)=1$. The plot shows the
    residual $A_{\rm recon}(k)-1$, where $A_{\rm recon}(k)$ denotes the
    reconstructed estimate of $A(k)$.
    }

    \label{fig:A_reconstruction}
\end{figure}

Figure~\ref{fig:A_reconstruction} presents the reconstructed correction factor $A(k)$ relative to the fiducial expectation, $A(k)=1$. Here, $A_{\rm recon}(k)$ denotes the reconstructed estimate of $A(k)$. The plot
shows the residual $A_{\rm recon}(k)-1$, with the dashed curve representing
the mean reconstructed residual and the shaded regions corresponding to
the $68\%$ and $95\%$ confidence intervals obtained from the Monte Carlo
realizations.

Across the reconstructed range, the mean value of $A(k)$ remains consistent with unity at $2\sigma$ confidence level, with no statistically significant scale dependence. The residual shown in the figure is centered around zero, indicating that the reconstructed correction factor remains consistent with the fiducial expectation, $A(k)=1$ within the $95\%$ ($2\sigma$) confidence region.
These results suggest that the enhancement observed in the reconstructed matter power spectrum is well accounted for by the nonlinear evolution
estimated by the PySCo simulations, whose nonlinear power spectrum is
consistent with the Mead2020 prescription over the relevant range. Within the statistical precision of the current CMB lensing measurements, the reconstructed correction factor remains consistent with unity, indicating no evidence for an additional scale-dependent modification beyond the fiducial nonlinear matter power spectrum.

Future CMB lensing measurements with improved sensitivity and a wider range of accessible scales will enable significantly tighter constraints on $A(k)$, providing a more stringent test of nonlinear structure formation and a useful framework for searching for possible departures from the standard cosmological model.

\section{Discussion}

The present reconstruction demonstrates that current CMB lensing measurements contain sufficient information to recover the broadband shape of the matter power spectrum over the range where the lensing kernel provides meaningful sensitivity. In the linear-input consistency test, the reconstructed residuals remain consistent with zero within the $68\%$ ($1\sigma$)
confidence region, indicating that the reconstruction does not introduce statistically significant spurious nonlinear features. The BAO null test similarly shows that, when a smooth no-BAO spectrum is used as input, the
reconstructed ratio $P(k)/P_{\rm smooth}(k)$ remains consistent with unity within the $68\%$ ($1\sigma$) confidence region, indicating that the
pipeline does not artificially generate BAO-like oscillations. For the observational reconstruction performed under the linear matter-power-
spectrum ansatz, the reconstructed spectrum begins to show an enhancement relative to the linear prediction at $k\gtrsim0.1\,{\rm Mpc}^{-1}$. The
enhancement reaches more than $3\sigma$ level where the reconstruction retains substantial sensitivity, while at higher wavenumbers the decreasing sensitivity of the lensing kernel increases the
uncertainties and reduces the apparent significance to approximately the $2\sigma$ level. Comparison with the nonlinear prediction shows that the reconstructed ratio follows the Mead2020 prediction up to $k\sim0.2\,{\rm Mpc}^{-1}$. The subsequent reconstruction of the
scale-dependent correction factor gives $A(k)$ consistent with unity at the $2\sigma$ ($95\%$) level over the reconstructed range, with no statistically significant scale dependence. Taken together, these results indicate that the observed high-$k$ enhancement is consistent with the standard nonlinear evolution of structure within the sensitivity of the current data.

The current analysis is subject to several specific limitations. The reconstruction is performed at fixed background cosmological parameters and uses the redshift evolution of the nonlinear matter power spectrum provided by the PySCo simulations. The broad projection of CMB lensing limits the effective resolution in wavenumber and makes the reconstruction progressively less constrained towards the boundaries of the reconstructed range, $10^{-3}\lesssim k\lesssim0.7\,{\rm Mpc}^{-1}$. The extraction of BAO-scale structure also depends on the choice of the smooth reference spectrum used to isolate the oscillatory component. Although the Limber approximation is adopted in the forward model, our numerical validation shows sub-percent agreement with the exact CAMB calculation over the multipole range relevant to this analysis, indicating that the Limber
approximation does not significantly affect the reconstructed results. 

These limitations motivate revisiting the reconstruction as higher-precision
CMB lensing data become available.
Future high-precision CMB lensing measurements will provide improved constraints on the matter power spectrum over a wider range of scales. In particular, measurements from experiments such as the Simons Observatory can provide substantially improved statistical precision and extended multipole coverage, allowing this reconstruction framework to be revisited
with tighter constraints. Such data could enable more precise tests of BAO-scale information and nonlinear structure formation, as well as more stringent constraints on possible scale-dependent departures through $A(k)$.

\acknowledgments
D.K.H. acknowledges financial support from the Indo-French Centre for the Promotion of Advanced Research (IFCPAR/CEFIPRA), New Delhi, India, through the Collaborative Scientific Research Programme (Project No. 6704-4, ``Testing flavors of the early universe beyond vanilla models with cosmological observations''); and from the Anusandhan National Research Foundation (ANRF), Government of India, under the ARG MATRICS scheme (Grant No. ANRF/ARGM/2025/000941/TS; project ``EPOCH: Exploring Primordial Origins in Cosmic Hierarchies'').
J.J. and A.S. acknowledge funding from the Korea Astronomy and Space Science Institute (KASI) through the project ``Research on the Principles of the Accelerating Expansion of the Universe'' (Project Code: 2026183201).

\appendix
\section{Numerical validation of the forward model}
\label{app:forward_model_validation}

The reconstruction pipeline developed in this work relies on the numerical evaluation of the forward model relating the matter power spectrum to the CMB lensing power spectrum. Before applying the reconstruction algorithm to either mock or observational data, it is therefore important to verify that the discretized implementation accurately reproduces the corresponding theoretical prediction.

To assess the convergence with respect to the number of reconstruction
nodes, we evaluate the forward model using three different choices,
$N_k=37$, $75$, and $150$, while keeping the fiducial cosmological model
and all other numerical settings fixed. For each choice of $N_k$, the
resulting CMB lensing power spectrum is compared with the corresponding
CAMB prediction through the percentage fractional difference. This allows us to
assess how the discretization affects the forward-model accuracy and
whether the adopted value of $N_k=75$ is sufficient.

\begin{figure}[H]
    \centering
    \includegraphics[width=0.5\linewidth]{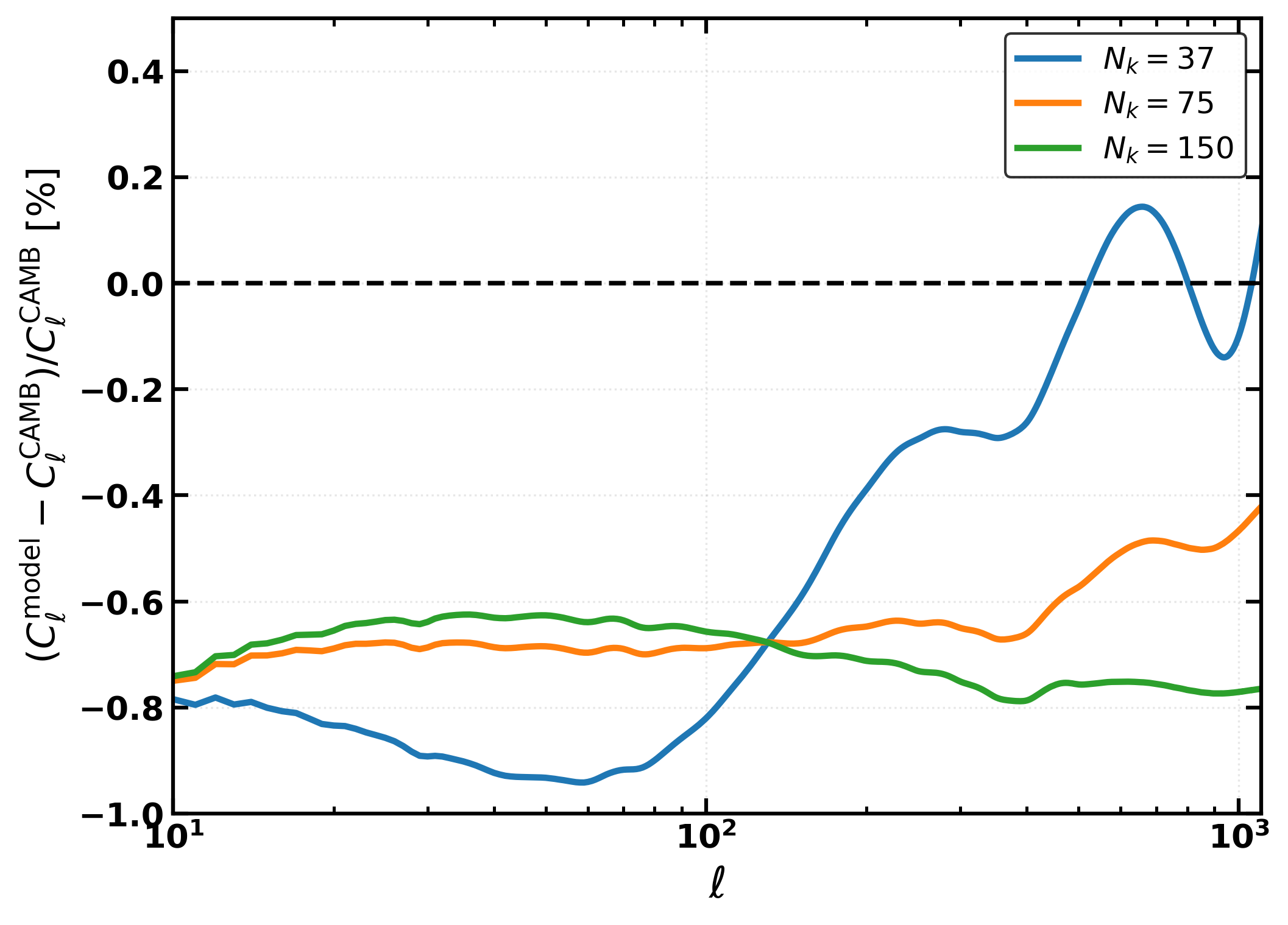}
    \caption{Percentage fractional difference between the discretized forward-model prediction
    and the corresponding CAMB lensing prediction for three choices of
    reconstruction nodes, $N_k=37$, $75$, and $150$. The comparison
    illustrates the convergence of the forward model with increasing
    $N_k$ and the saturation of the discretization error around the adopted
    value $N_k=75$. The fractional difference is defined as
    $100\times(C_\ell^{\rm model}-C_\ell^{\rm CAMB})/C_\ell^{\rm CAMB}$.
    }
    \label{fig:forward_validation}
\end{figure}

The forward-model predictions converge as the number of reconstruction
nodes is increased. In particular, the results for $N_k=75$ and
$N_k=150$ become nearly indistinguishable over the multipole range
relevant to the analysis, indicating that the discretization has
effectively saturated by $N_k=75$. We therefore adopt $N_k=75$ for the
reconstruction presented in this work.

\bibliographystyle{JHEP}
\bibliography{references}

\end{document}